\documentclass[twocolumn]{emulateapj} 

\usepackage{amsfonts}
\usepackage{amsbsy}
\usepackage{natbib}
\usepackage{graphicx}
\usepackage{ulem}

\newcommand{\dd}{\mathrm{d}} %differential for integrals
\newcommand{\myvec}[1]{\boldsymbol{#1}}
\newcommand{\mymat}[1]{#1}

\slugcomment{\textit{Draft Version, Nov. 22, 2010. Accepted to ApJ} }
 
\alph{footnote}

\begin{document}

\title{3D Reconstruction of the Density Field: An SVD Approach to Weak Lensing Tomography}

\author{J. T. VanderPlas and A. J. Connolly}
\affil{Astronomy Department, University of Washington, Box 351580, 
  Seattle, WA 98195-1580}
\and
\author{B. Jain and M. Jarvis}
\affil{Department of Physics and Astronomy, University of Pennsylvania, 
  209 South 33rd Street, Philadelphia, PA 19104-6396}

%\author{Jake VanderPlas\altaffilmark{1}, Andrew Connolly\altaffilmark{1}, 
%  Bhuvnesh Jain\altaffilmark{2}, and Mike Jarvis\altaffilmark{2}}
%\noaffiliation

%\altaffiltext{1}{Astronomy Department, University of Washington, 
%  Box 351580, Seattle, WA 98195-1580}
%\altaffiltext{2}{Department of Physics and Astronomy, 
%  University of Pennsylvania, 
%  209 South 33rd Street, Philadelphia, PA 19104-6396}

%\date{\today}		 		% your own text, a date, or \today

% --------------------- end of the preamble ---------------------------
			% REQUIRED
\bibliographystyle{apj}

\begin{abstract}
  We present a new method for constructing three-dimensional
  mass maps from gravitational lensing shear data.  We solve the lensing
  inversion problem using truncation of singular values 
  (within the context of generalized
  least squares estimation) without a priori assumptions about the
  statistical nature of the signal.   This singular value framework 
  allows a quantitative
  comparison between different filtering methods: we evaluate our method
  beside the previously explored Wiener filter approaches.
  Our method yields near-optimal angular resolution of the lensing
  reconstruction and allows cluster sized halos to be de-blended robustly.
  It allows for mass reconstructions which are
  2-3 orders-of-magnitude faster than the Wiener
  filter approach; in particular, we estimate that an all-sky
  reconstruction with arcminute resolution could be performed
  on a time-scale of hours. We find however that linear, 
  non-parametric reconstructions have a fundamental limitation in the
  resolution achieved in the redshift direction.
\end{abstract}


\section{Introduction}

In recent years, the study of gravitational lensing has proved a
valuable tool in advancing our understanding of the universe.  Because
deflection of light in a gravitational field is a well-understood
aspect of Einstein's General Relativity, it offers a unique method of
mapping the mass distribution within the universe (including the dark
matter), which is free of any astrophysical bias.  Though much insight
can be gained from the two dimensional projection of the matter distribution
\citep[see, e.g.][]{Clowe06}, a nonparametric technique that can map
the full 3D matter distribution is in principle attainable. 

\citet{Taylor01}, \citet[][hereafter HK02]{Hu02} and 
\citet{Bacon03} first looked at
non-parametric 3D mapping of a gravitational potential.  HK02
presented a linear-algebraic method for \textit{tomographic mapping}
of the matter distribution -- splitting the sources and lenses into
discrete planes in redshift.  They found that the inversion along each
line-of-sight is ill-conditioned, and requires regularization through 
\textit{Wiener filtering}.  Wiener filtering reduces reconstruction noise
by using the expected statistical properties of the signal as a prior: 
for the present problem, this prior is the nonlinear mass power spectrum.  
\citet[][hereafter STH09]{Simon09}
made important advances to this method by constructing an efficient
framework in which the inversions for every line-of-sight are computed
simultaneously, allowing for greater flexibility in the
type of filter used.  They introduced two types of Wiener filters: 
a ``radial Wiener filter'', based on the HK02 method, 
and a ``transverse Wiener filter'', 
based on the Limber approximation to
the 3D mass power spectrum.
They showed that the use of a generalized form of either
filter leads to a biased result -- the filtered reconstruction of the
line-of-sight matter distribution for a localized lensing mass is both
shifted and spread-out in redshift.

One issue with the Wiener filter approach is the assumption of
Gaussian statistics in the reconstructed signal.  In reality, the matter 
distribution at relevant scales can be highly non-Gaussian.  
It is possible that the redshift bias found in STH09 is not inherent
to nonparametric linear mapping, but rather a result of this deficiency
in the Wiener filtering method.

In this work, we develop an alternate noise-suppression scheme for
tomographic mapping that,  unlike Wiener filtering, has no dependence
on assumptions about the signal.   Our goal is to explore improvements
in the reconstruction and examine, in particular, the recovery of
redshift information using the different methods. We begin in 
Section~\ref{Method} 
by discussing the tomographic weak-lensing model developed by HK02
and STH09 and presenting our estimator for the density parameter, $\delta$.  
In Sections~\ref{Results} and \ref{Conclusions} we implement this method
for a simple case, and compare the results with those of the STH09 transverse
and radial Wiener filters.

\section{Method}
\label{Method}

For tomographic weak lensing, we are concerned with three quantities:  the 
complex-valued shear $\gamma(\vec\theta,z)$, the real-valued convergence
$\kappa(\vec\theta,z)$, and the dimensionless density parameter 
$\delta(\vec\theta,z)$.  
The relationship between $\gamma$ and $\kappa$ is given
by a convolution over all angles $\vec\theta$, and the density $\delta$ is
related to $\kappa$ by a line-of-sight integral over the lensing efficiency
function, $W(z,z_s)$.  The key observation is that in the weak lensing regime,
each of these operations is linear: if the variables are discretized, 
they become systems of linear equations, 
which can in principle be solved using standard matrix methods.

\subsection{Linear Mapping}
\label{LinearMapping}

To achieve this, we create a common pixel binning of 
$N_x$ by $N_y$ equally sized square pixels
of angular width $\Delta\theta_x = \Delta\theta_y$.  
Within each of the $N_x N_y \equiv N_{xy}$ individual
lines of sight, we bin $\gamma$ into $N_s$ source-planes,
and bin $\delta$ into $N_l$ lens-planes, $N_l \le N_s$.  
Thus we have two 1D data vectors, which are concatenations of
the line-of-sight vectors within each pixel:
 $\myvec\gamma$, of length $N_{xy} N_s$; 
and $\myvec\delta$, of length $N_{xy} N_l$.  
(Note that throughout this section, boldface denotes a vector quantity.)  
As a result of this binning, we can write the discretized lensing 
equations in a particularly simple form:
\begin{equation}
  \label{M_gd}
  \myvec\gamma = M_{\gamma\delta}\myvec\delta + \myvec{n_\gamma}
\end{equation}
where $\myvec\gamma$ is the vector of binned shear observations with noise
given by $\myvec{n_\gamma}$, and $\myvec\delta$ is the vector 
of binned density parameter.  For details on the form of 
the matrix $M_{\gamma\delta}$, refer to Appendix~\ref{app}.

The linear estimator $\myvec{\hat{\delta}}$ of the signal is found by 
minimizing the quantity
\begin{equation}
  \chi^2 = (\myvec\gamma-\mymat{M_{\gamma\delta}}\myvec{\delta})^\dagger 
  \mymat{\mathcal{N}_{\gamma\gamma}}^{-1} (\myvec\gamma-\mymat{M_{\gamma\delta}}\myvec{\delta})
\end{equation}
where $\dagger$ indicates the conjugate transpose, and 
$\mymat{\mathcal{N}_{\gamma\gamma}} \equiv 
\langle \myvec{n_\gamma}\myvec{n_\gamma}^\dagger\rangle$
is the noise covariance of the measurement $\myvec{\gamma}$,
and we assume $\langle \myvec{n_\gamma}\rangle = 0$.  The best linear
unbiased estimator for this case is due to \citet{Aitken34}: 
\begin{equation}
  \label{Aitken_estimator}
  \myvec{\hat{\delta}_A} \equiv
  \left[\mymat{M_{\gamma\delta}}^\dagger
    \mymat{\mathcal{N}_{\gamma\gamma}}^{-1}\mymat{M_{\gamma\delta}}\right]^{-1}
  \mymat{M_{\gamma\delta}}^\dagger
  \mymat{\mathcal{N}_{\gamma\gamma}}^{-1}\myvec{\gamma}
\end{equation}
The noise properties of this estimator can be made clear by
defining the matrix
$\mymat{\widetilde{M_{\gamma\delta}}}\equiv 
\mymat{\mathcal{N}_{\gamma\gamma}}^{-1/2}\mymat{M_{\gamma\delta}}$ and
computing the singular value decomposition (SVD)
$\mymat{\widetilde{M_{\gamma\delta}}} 
\equiv \mymat{U}\mymat{\Sigma}\mymat{V}^\dagger$.
Here $\mymat{U}^\dagger \mymat{U} = \mymat{V}^\dagger \mymat{V} = \mymat{I}$ 
and $\Sigma$ is the square diagonal matrix of singular values 
$\sigma_i\equiv\Sigma_{ii}$, ordered such that $\sigma_i\ge\sigma_{i+1}$, 
$i\ge 1$.  Using these properties, the Aitken estimator can be equivalently 
written
\begin{equation}
  \label{Aitken_SVD}
  \myvec{\hat{\delta}_A} = \mymat{V} \mymat{\Sigma}^{-1} 
  \mymat{U}^\dagger \mymat{\mathcal{N}_{\gamma\gamma}}^{-1/2} \myvec{\gamma}
\end{equation}
It is apparent in this expression that the presence of small singular values 
$\sigma_i \ll \sigma_1$ can lead to extremely large diagonal entries 
in the matrix $\mymat{\Sigma^{-1}}$, which in turn amplify the errors in the
estimator $\myvec{\hat{\delta}_A}$. 
This can be seen formally by expressing the noise covariance
in terms of the components of the SVD:
\begin{equation}
  \label{Ndd_decomp}
  \mymat{\mathcal{N}_{\delta\delta}} = 
  \mymat{V}\mymat{\Sigma}^{-2}\mymat{V}^\dagger.
\end{equation}
The columns of the matrix $\mymat{V}$ are eigenvectors of
$\mymat{\mathcal{N}_{\delta\delta}}$, with eigenvalues $\sigma_i^{-2}$.
When many small singular values are present, the noise will dominate
the reconstruction, and it is necessary to 
use a more sophisticated estimator to recover the signal.

\subsection{SVD Filtering}
\label{sing_val_formalism}
One strategy that can be used to reduce this noise is to add a penalty function
to the $\chi^2$ that will suppress the large spikes in signal.  This
is the Wiener filter approach explored by HK02 and STH09.
A more direct noise-reduction method, which does not require knowledge 
of the statistical properties of the signal, involves approximating 
the SVD in Equation~\ref{Aitken_SVD} to remove the contribution of the 
high-noise modes. We choose a cutoff value
$\sigma_{\rm cut}$, and determine $n$ such that 
$\sigma_n > \sigma_{\rm cut} \ge \sigma_{n+1}$.
We then define the truncated matrices 
$\mymat{U_n}$, $\mymat{\Sigma_n}$, and $\mymat{V_n}$,
such that $\mymat{U_n}$ ($\mymat{V_n}$) contains the first $n$ columns of 
$\mymat{U}$ ($\mymat{V}$),
and $\mymat{\Sigma}_n$ is a diagonal matrix of the largest $n$ singular 
values, $n \le n_{\rm max}$.
To the extent that $\sigma_{\rm cut}^2 \ll \sum_{i=1}^n\sigma_i^2$, 
the truncated matrices satisfy
\begin{equation}
  \mymat{U_n}\mymat{\Sigma_n}\mymat{V_n}^\dagger \approx 
  \mymat{U}\mymat{\Sigma}\mymat{V}^\dagger = \mymat{\widetilde{M_{\gamma\delta}}}
\end{equation}
and the signal estimator in Equation~\ref{Aitken_estimator} can be approximated
by the SVD estimator:
\begin{equation}
  \label{SVD_estimator}
  \myvec{\hat{\delta}}_{\rm svd}(n) \equiv 
  \mymat{V}_n \mymat{\Sigma}_n^{-1} 
  \mymat{U}_n^\dagger \mymat{\mathcal{N}_{\gamma\gamma}}^{-1/2} \myvec{\gamma}
\end{equation}
This approximation is optimal in the sense that it preferentially
eliminates high-noise orthogonal components in $\myvec{\delta}$ 
(cf.\ equation~\ref{Ndd_decomp}), leading to an estimator which is much
more robust to noise in $\myvec{\gamma}$.

SVDs are often used in the context of Principal Component Analysis, 
where the square of the singular value is equal to the variance described 
by the corresponding principal component.  The variance can be
thought of, roughly, as a measure of the information contributed by the
vector to the matrix in question.  It will be useful for
us to think about SVD truncation in this way.  To that end, 
we define a measure of the truncated variance for a given value of $n$:
\begin{equation}
  \label{v_cut}
  v_{\rm cut}(n) = 1 - \frac{\sum_{i=1}^n\sigma_i^2}
  {\sum_{i=1}^{n_{\rm max}}\sigma_i^2}
\end{equation}
such that $0\le v_{\rm cut}\le 1$.  If $v_{\rm cut} = 0$, then $n = n_{\rm max}$ and we are 
using the full Aitken estimator.  As $v_{\rm cut} \rightarrow 1$, we are increasing the amount
of truncation.

In practice, taking the SVD of the transformation matrix 
$\mathcal{N}_{\gamma\gamma}^{-1/2}M_{\gamma\delta}$
is not entirely straightforward: the matrix is of size 
$(N_{xy}N_s) \times (N_{xy}N_l)$.
With a $128 \times 128$-pixel field, 20 lens-planes, and 25 source-planes,
the matrix contains $1.3\times 10^{11}$ mostly nonzero complex entries, 
amounting to 2TB in memory (double precision).  
Computing the SVD for a non-sparse matrix of this size is far from trivial.

We have developed a technique to speed-up this process, which involves decomposing
the matrices $\mathcal{N}_{\gamma\gamma}$ and $M_{\gamma\delta}$ into tensor
products, so that the full SVD can be determined through computing SVDs of
two smaller matrices: an $N_s\times N_l$ matrix, 
and an $N_{xy}\times N_{xy}$ matrix.
The second of these individual SVDs can be approximated using the
Fourier-space properties of the $\gamma\to\kappa$ mapping.  The result is
that the entire SVD estimator can be computed very quickly.  The details
of this method are described in Appendix~\ref{app}.

\begin{figure*}[t] 
 \centering
 \plotone{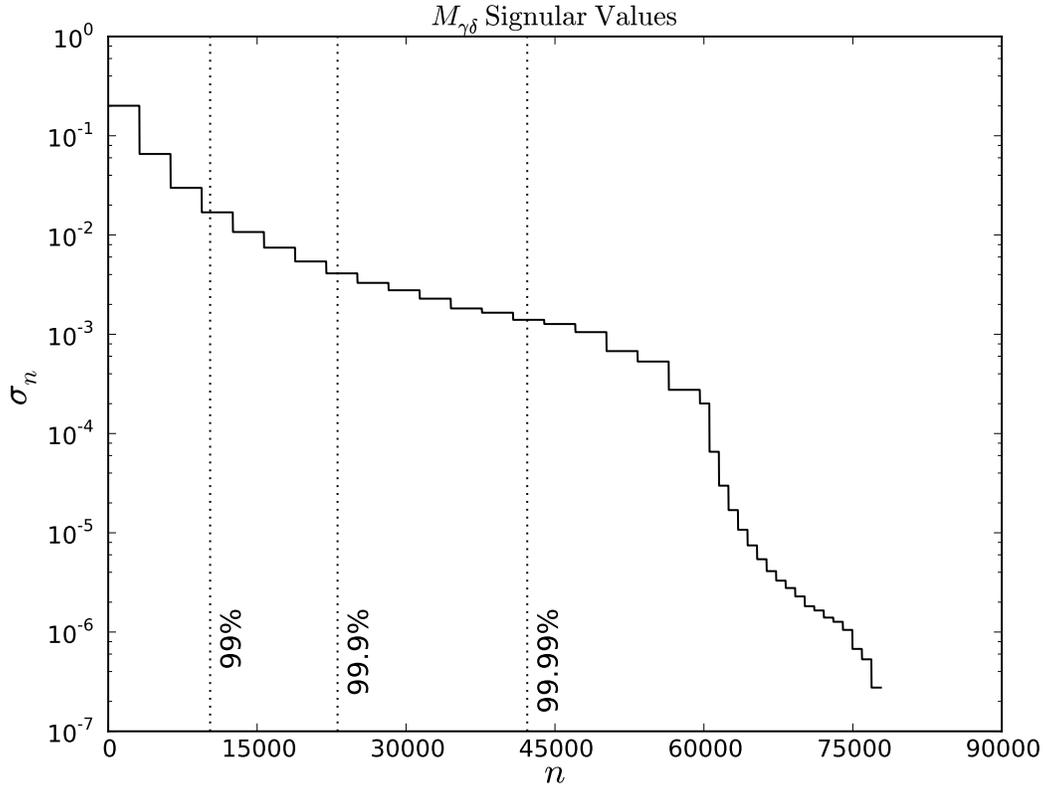}
 \caption{
   Ordered singular values of the matrix
   $\mymat{\widetilde{M_{\gamma\delta}}}$.
   The dotted lines show the values of $n$ 
   such that 99\%, 99.9\%, and 99.99\% of the variance is preserved.
   The sharp drop-off near $n=60,000$ is due to the $10^{-3}$ 
   deweighting of border pixels.
   \label{fig_sing_vals}}
\end{figure*}

\section{Results}
\label{Results}
\label{Parameters}
Using the above formalism, we can now explore the
tomographic weak lensing problem using the 
techniques of Section~\ref{Method}.
For the following discussion, we will use a field of approximately 
one square degree: a $64 \times 64$ grid of
$1^\prime \times 1^\prime$ pixels, with 25 source redshift
bins ($0\le z\le 2.0$, $\Delta z = 0.08$) and 20 lens redshift bins
($0\le z\le 2.0$, $\Delta z = 0.1$).  This binning approximates the 
expected photometric redshift errors of future surveys.
We suppress edge effects by increasing
the noise of all pixels within $4^\prime$ of the field border
by a factor of $10^3$, effectively
deweighting the signal in these pixels (cf.\ STH09).  The noise
for each redshift bin is set to $n_i = \sigma_\gamma/\sqrt{N_i}$, where 
$\sigma_\gamma$ is the intrinsic ellipticity dispersion, and
$N_i$ is the number of galaxies in the bin.  We assume $\sigma_\gamma = 0.3$,
and 70 galaxies per square arcminute, with a redshift distribution given by
\begin{equation}
  \label{gal_z_dist}
  n(z) \propto z^2\exp{\left[-(z/z_0)^{3/2}\right]},
\end{equation}
with $z_0 = 0.57$.  We assume a flat cosmology with 
$h=0.7$, $\Omega_M = 0.27$  and $\Omega_\Lambda = 0.73$ at the present day.

\subsection{Singular Values}
The singular values of the transformation matrix for this configuration 
are depicted in Figure~\ref{fig_sing_vals}.  The step pattern visible in
this plot is due to the fact that the noise across 
each source plane is identical,
aside from the $4^\prime$ deweighted border.  
It is apparent from this figure that the large
majority of the singular values are very small: 99.9\% of the variance
in the transformation is contained in less than $1/3$
of the singular values.  The large number of very small
singular values will, therefore, dominate in the Aitken estimator 
(Equation~\ref{Aitken_SVD}), leading to the very noisy unfiltered 
results seen in HK02.  

\subsection{Evaluation of the SVD Estimator}
\label{SVD_Evaluation}
To evaluate the performance of the SVD filter, we first create a
field-of-view containing a single halo at redshift $z = 0.6$.
One well-supported parametrization of halo shapes is the
NFW profile \citep{NFW97}.  We use the analytic form of the
shear and projected density due to an NFW profile, given by equations
13-18 in
\citet{Takada03}.

\begin{figure*}[p]
 \centering
 \plotone{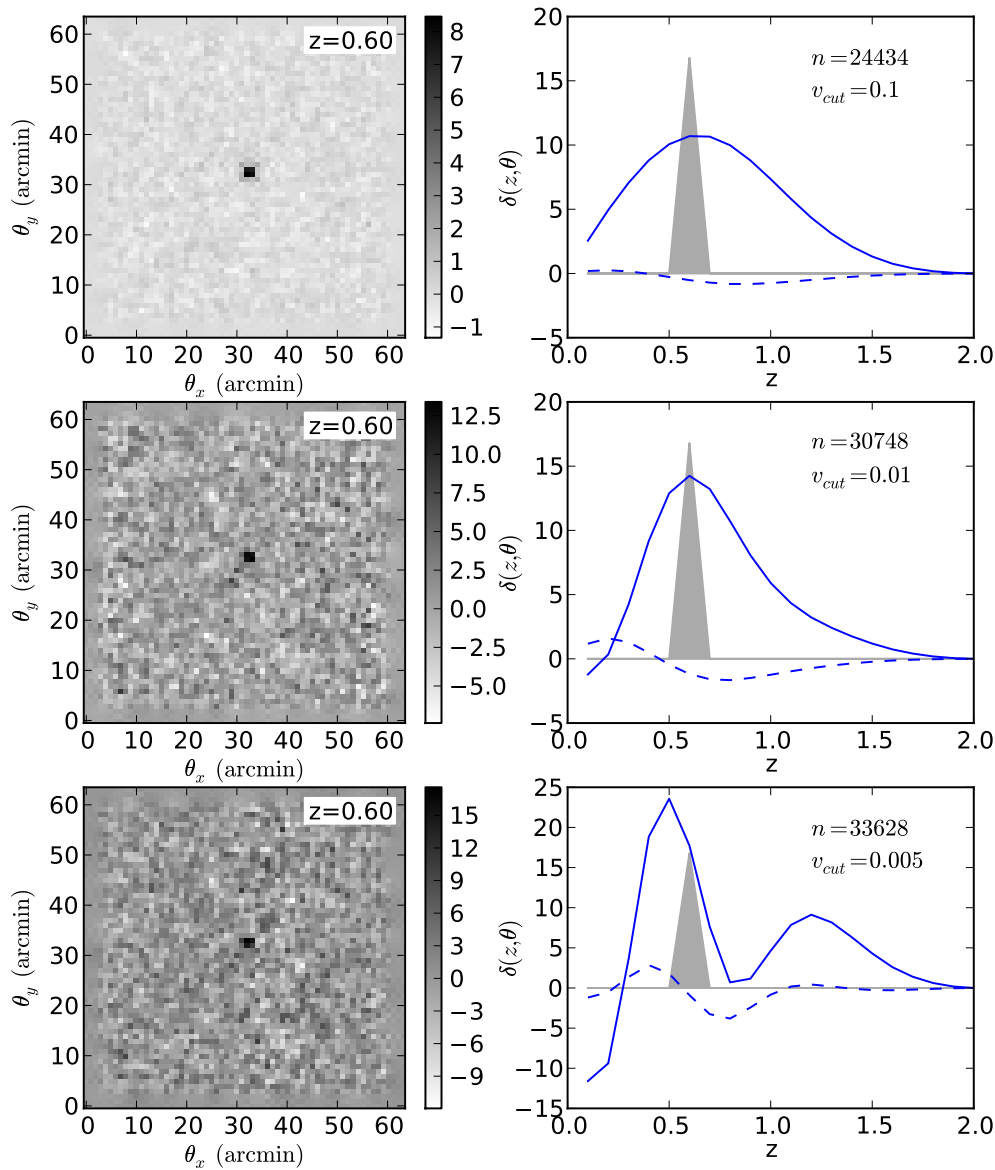}
 \caption{
   The effect of SVD truncation on a single $z=0.6$ NFW halo
   in the center of the field, for three different levels of filtering.
   \textit{left column:}
   reconstructed density parameter $\delta(\theta)$ in the $z=0.6$ lens-plane.
   The true matter distribution is represented by a tight ``dot'' in the 
   center of the plot.
   \textit{right column:}
   line-of-sight profile at the central pixel.  The grey shaded area 
   shows the input density parameter.  The solid line shows the E-mode signal, 
   while the dashed line shows the B-mode signal.
   $n$ gives the number of singular values
   used in the reconstruction (out of a total $n_{\rm max}=81920$), 
   and $v_{\rm cut}$ gives the amount of variance cut by the truncation 
   (Equation~\ref{v_cut}); the level of filtration decreases from the top
   panels to the bottom panels.  
   The bottom panels show a case of under-filtering:
   for small enough $v_{\rm cut}$, the noise overwhelms the signal and creates
   spurious peaks along the line-of-sight.
   \label{fig_los_plot}} 
\end{figure*}

\begin{figure*}[t]
 \centering
 \plotone{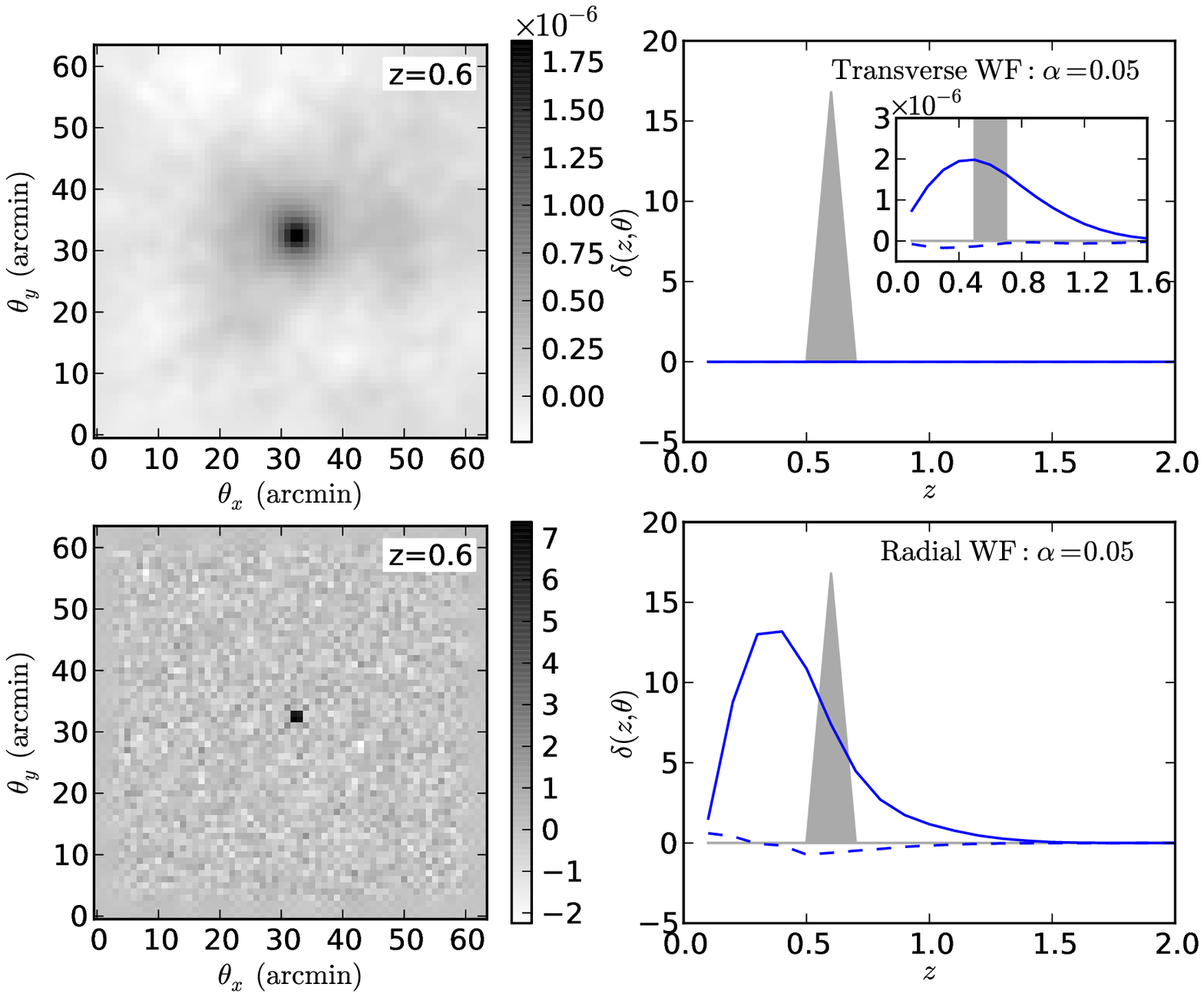}
 \caption{
   The effect of Wiener filtering on the same input as 
   Figure~\ref{fig_los_plot}. Here we have used both transverse 
   \textit{(top panels)}
   and radial \textit{(bottom panels)} Wiener filtering, both down-tuned 
   by $\alpha = 0.05$ (the value recommended by STH09).  
   The transverse Wiener filter suppresses the response by several 
   orders of magnitude; a closer view of the line-of-sight peak is shown 
   in the inset plot.  The radial Wiener filter gives similar angular 
   results to the SVD filter, but takes much longer to compute.
   \label{fig_los_plot_ST}} 
\end{figure*}

We reconstruct the density map using the SVD filter 
(Figure~\ref{fig_los_plot}) with the above survey parameters.  
We show the results for three different values of $v_{\rm cut}$: 
0.1, 0.01, and 0.005.  In all three cases, the halo is easily 
detected at its correct location (left panels),
although as $v_{\rm cut}$ decreases, there is more noise in the 
surrounding field.  The right panels show the computed density profile
along the line of sight for the central pixel. 
The peak of this curve is reasonably close to the correct redshift, 
but there is a significant spread in redshift, as well as a bias.
As the level of SVD filtering (measured by $v_{\rm cut}$) decreases, 
the magnitude of these effects decreases, but the increased noise 
leads to spurious peaks. 

Similar plots for the transverse Wiener filter recommended by STH09 are
shown in the upper panels of Figure~\ref{fig_los_plot_ST}, using their 
recommended value of $\alpha = 0.05$. 
The response shows a significant spread in angular space, and 
the signal is seen to be suppressed by six orders-of-magnitude along with
a similar suppression of the noise. 
These effects worsen, 
in general, as the filtering level $\alpha$ increases.
Mathematically it is apparent why the transverse filter performs so poorly:
the small singular values primarily come from the line-of-sight 
part of the mapping, and the this filter has no effect along the line-of-sight.

The effect of the radial Wiener filter is shown in the bottom panels of 
Figure~\ref{fig_los_plot_ST}.
It shares the positive aspects of the SVD filter, having very 
little signal suppression or angular spread.
However, this filter uses some priors on the statistical form of 
the signal that are not as physically well-motivated as those for the 
transverse Wiener filter.  In contrast, the SVD filter does not make 
any prior assumptions about the signal. In this way, the SVD reconstruction
can be thought of as even more non-parametric than the Wiener filter
reconstructions.

\subsection{Comparison of Estimators}
\label{Comparison}

\begin{figure*}[t]
 \centering
 \plotone{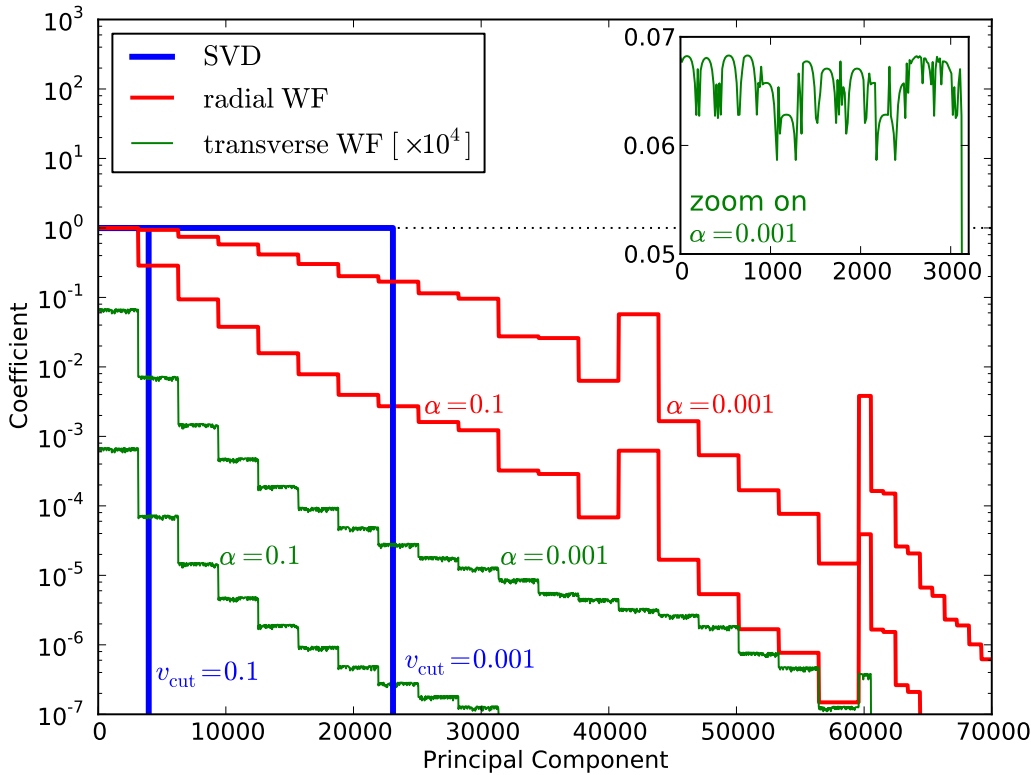} 
 \caption{Contribution of each shear mode to the reconstruction
 for three different filters.  The dotted line at $10^0$ represents the
 unfiltered result.  Each filtering method leads to a different
 weighting of the shear modes.  The SVD filter, by design, completely 
 removes higher-order modes beyond a given cutoff, while the Wiener 
 filter deweights modes in a more gradual fashion.
 Note that the transverse Weiner filter deweights
 all modes by up to seven orders of magnitude; it has been scaled by
 a factor of $10^4$ for this plot.  The inset plot shows a
 closeup of the fluctuations within each ``step'' of the transverse
 filter.  These fluctuations lead to angular spread in the response
 (see discussion in Section~\ref{Comparison})
 \label{fig_mode_comparison} }
\end{figure*}

The SVD framework laid out in Section~\ref{sing_val_formalism} can be used
to quantitatively compare the behavior of different estimators.  A general
linear estimator has the form
\begin{equation}
  \myvec{\hat\delta_R} = \mymat{R}\myvec\gamma
\end{equation}
for some matrix $\mymat{R}$.
This general estimator can be expressed in terms of the components of
the unbiased estimator (Equation~\ref{Aitken_SVD}):
\begin{equation}
  \mymat{R} = \mymat{V_R} \mymat{\Sigma}^{-1} 
  \mymat{U}^\dagger \mymat{\mathcal{N}_{\gamma\gamma}}^{-1/2}.
\end{equation}
Here the matrices $\mymat\Sigma$, $\mymat{U}$ and 
$\mymat{\mathcal{N}_{\gamma\gamma}}$ are defined as in 
Equation~\ref{Aitken_SVD}, and we have defined the matrix
\begin{equation}
  \mymat{V_R} \equiv \mymat{R}\mymat{\mathcal{N}_{\gamma\gamma}}^{1/2}
  \mymat{U}\mymat{\Sigma}
\end{equation}
The rows of the matrix $\mymat{\Sigma}^{-1} \mymat{U}^\dagger 
\mymat{\mathcal{N}_{\gamma\gamma}}^{-1/2}$ provide a convenient basis
in which to work: they are the weighted principal components of the shear, 
ordered with decreasing signal to noise.  The norm of
the $i^{\rm th}$ column of $\mymat{V_R}$ measures the contribution of
the $i^{\rm th}$ mode to the reconstruction of $\delta$.  For the unfiltered
estimator, $\mymat{V_R}=\mymat{V}$ and all the norms are unity.
This leads to a very intuitive comparison between different filtering schemes.
Figure~\ref{fig_mode_comparison} compares the column-norms of $V_R$ 
for the SVD filter with those of the radial and
transverse Wiener filters.
  
The steps visible in the plot originate the same way as the steps in
Figure~\ref{fig_sing_vals}: the flatness of each step comes from the 
assumption of uniform noise in each source plane.  This plot shows 
the tradeoff between noise and bias.  The flat line at norm=$10^0$ 
represents a noisy but unbiased estimator.  Any departure
from this will impose a bias, but can increase signal-to-noise.
There are two important observations from this figure.
First, because each step on the plot is relatively flat for
the SVD filter and radial Wiener filter, we don't expect much bias 
\textit{within} each lens plane.  
The transverse filter, on the other hand, has fluctuations
at the $10\%$ level within each step (visible in the inset of 
Figure~\ref{fig_mode_comparison}), which will lead to a noticeable bias
within each lens plane, resulting in the degraded angular resolution
of the reconstruction seen in Figure~\ref{fig_los_plot_ST}.
Second, the transverse Wiener filter deweights even the highest 
signal-to-noise modes by many orders of magnitude, resulting 
in the signal suppression seen in Figure~\ref{fig_los_plot_ST}.
The SVD filter and radial Wiener filter, on the other hand,
have weights near unity for the highest signal-to-noise modes.
These two observations show why the SVD filter and radial Wiener filter 
are the more successful noise reduction techniques for the present problem.

\subsection{Noise Properties of Line-of-Sight Modes}
\label{SN_modes}
As seen in equation \ref{Ndd_decomp}, the columns of $\mymat{V}$ provide
a natural orthogonal basis in which to express the signal $\myvec{\delta}$.  
It should be emphasized that this eigenbasis is 
valid for any linear filtering scheme: the untruncated
SVD is simply an equivalent re-expression of the original transformation.
Examining the characteristics of these eigenmodes can yield insight
regardless of the filtering method used.

\begin{figure*}[t]
 \centering
 \plottwo{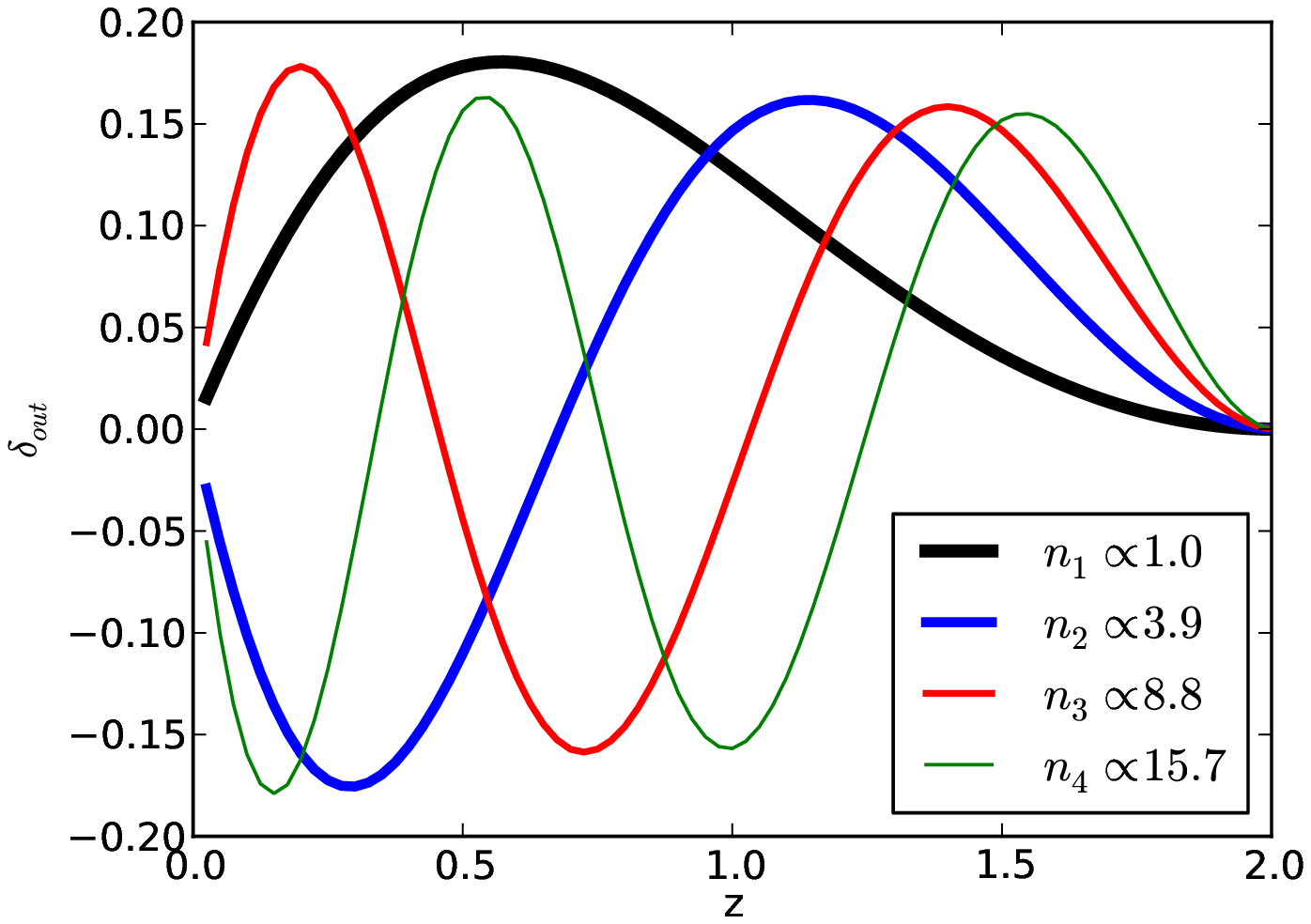}{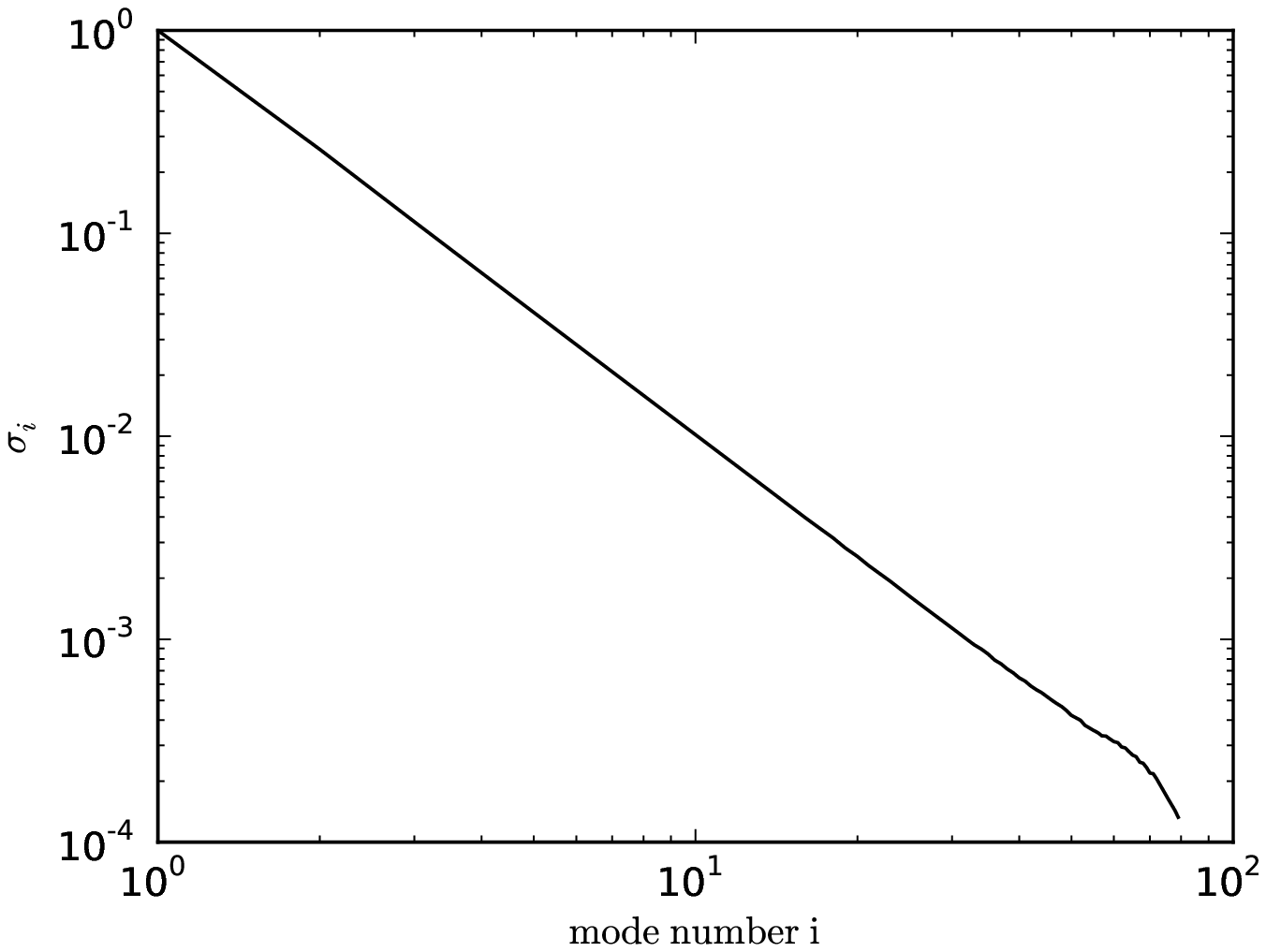}  
 \caption{\textit{left panel:} 
   The radial components of the first four
   columns of the matrix $\mymat{V}$ (see section~\ref{LinearMapping}).
   This is calculated for 100 equally spaced redshift bins $(0\le z\le 2.5)$ 
   in $\myvec{\gamma}$, and 80 bins $(0\le z\le 2.0)$ in $\myvec{\delta}$
   These orthogonal eigenmodes are analogous to radial Fourier modes.  Each
   is labeled by its relative noise level, $n_i = (\sigma_i/\sigma_1)^{-1}$.
   \textit{right panel:}
   The singular values $\sigma_i$ associated with the 80 radial eigenmodes.
 \label{fig_radial_modes} }
\end{figure*} 

The radial components of the first four eigenmodes are plotted in 
figure~\ref{fig_radial_modes}.  Each is labeled by its normalized 
noise level, $n_i \equiv (\sigma_i/\sigma_1)^{-1}$.
The total number of modes will be equal to the number of output redshift 
bins; here, for clarity, we've used 80 equally-spaced bins out to redshift 2.0.
As the resolution is lessened, the overall shape and relative noise level of 
the lower-order modes is maintained.
These radial modes are analogous to angular Fourier modes,
and are related to the signal-to-noise KL modes discussed in HK02.
It is clear from this plot that any linear, non-parametric
estimator will be fundamentally limited in its redshift resolution: 
the noise level of the $i^{th}$ mode approximately scales as
\begin{equation}
  n_i\ \widetilde{\propto}\ i^2
\end{equation}
The signal-to-noise level for any particular halo will depend on
its mass and redshift. The magnitude of the signal
scales linearly with mass (see discussion in STH09), 
but the redshift dependence is more complicated: it is
affected by the lensing efficiency function, which depends on the redshift
of the lensed galaxies.  Using the above survey parameters, with an NFW halo 
of mass $M_{200} = 10^{15} \rm{M_\odot}$ and redshift $z=0.6$, 
the signal-to-noise ratio 
of the central pixel for the fundamental radial mode is $\sim 5.9$,
consistent with the results for Wiener filtered reconstructions of
singular isothermal halos explored in STH09.
This means that for even the largest halos, with a very deep survey,
only the first few modes will contribute significantly 
to the reconstructed halo.  Adding higher-order modes can in theory 
provide redshift information, but at the cost of increasingly high 
noise contamination. This is a general result which will apply to all 
nonparametric linear reconstruction algorithms.

This lack of information in the redshift direction leads directly 
to an inability to accurately determine halo masses:
the lensing equations relate observed shear $\gamma$ 
to density parameter $\delta$, which is related to mass in a
redshift-dependent way.  This is a fundamental limitation on the
ability of linear nonparametric methods to determine halo masses
from shear data. Indeed, even moving to fully parametric models, 
line-of-sight effects can lead to halo mass errors of 
20\% or more \citep{Hoekstra03, dePutter05}.

\subsection{Reconstruction of a Realistic Field}
\label{Realistic_Field}

\begin{figure*}%[t] 
 \centering
 \plotone{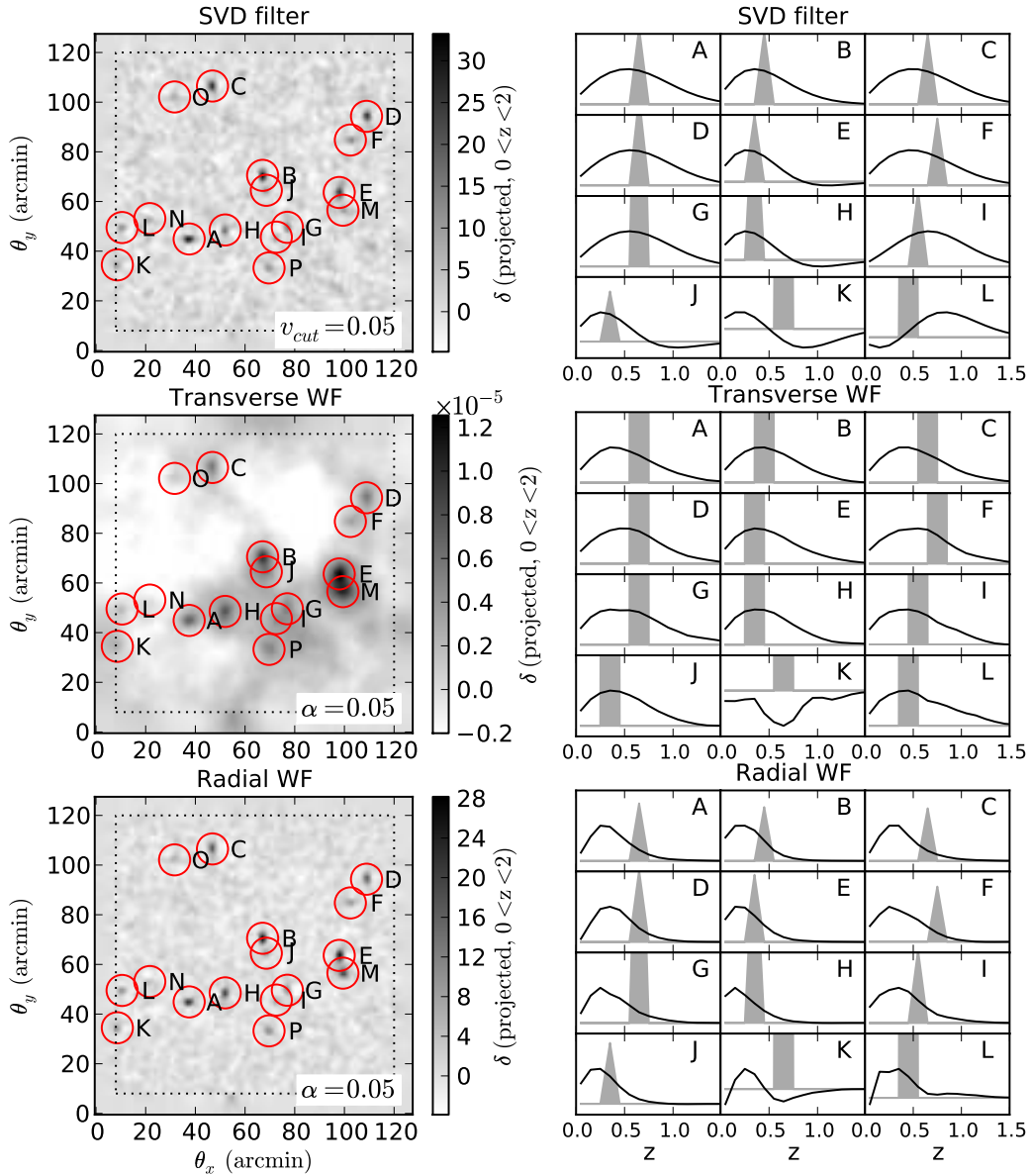} 
 \caption{
   Reconstruction of an artificial shear field with the 
   SVD filter \textit{(top panels)}, 
   Transverse Wiener filter \textit{(middle panels)}, 
   and Radial Wiener filter \textit{(bottom panels)}.  
   The left column shows the projected density reconstruction
   across the field using each method, all
   smoothed with a 1-pixel wide Gaussian filter.  
   Red circles indicate the true locations of the input halos.
   The right column shows the line-of-sight distributions of the
   twelve most massive NFW halos, labeled A-L.  The masses and redshifts of
   the halos are listed in Table~\ref{halo_table}.
   The signal suppression of the transverse Wiener filter seen in 
   Figure~\ref{fig_los_plot_ST} is apparent in the color-bar scaling
   of the middle panels. The anomalous results seen in halo K are due to
   its proximity to the deweighted border.  As suggested by the discussion
   in Section~\ref{SN_modes}, none of the three methods succeed 
   in recovering precise redshifts of the halos.
  \label{fig_many_halos} }
\end{figure*}

To compare the performance of the three filtering methods 
for a realistic field, we
create a 4 square degree field with approximately 20 halos between masses of
$2\times 10^{14}$ and $8\times 10^{14} {\rm M_\odot}$ with a mass 
distribution approximating the cluster mass function of \citet{Rines07},
and a redshift distribution given by Equation~\ref{gal_z_dist},
adding a hard cutoff at $z=1.0$.  
These parameters are chosen to approximate the true 
distribution of observable halos in a field this size.
The results of the reconstruction are shown in 
Figure~\ref{fig_many_halos} 

The red circles are the locations of the input halos, not the result 
of some halo-detection algorithm.  However, it is clear that, 
for at least most of the mass range, we are able to produce a map 
for which any reasonable detection algorithm should detect the halos
in the correct locations.  A few of the lower mass halos would 
certainly be missed though, since they are not significantly 
different from the noise peaks in the image.

In practice, one may vary the parameter $v_{\rm cut}$ as in 
Figure~\ref{fig_los_plot}
to trade-off robustness of detecting peaks with resolution in angle and in 
redshift. As shown in Section~\ref{Comparison}, we expect
filtering to introduce very little bias in angular resolution, 
so large values of $v_{\rm cut}$ lead to the most robust angular results. 
On the other hand, as shown in Section~\ref{SN_modes},
filtering introduces an extreme bias along the line-of-sight.
The effects of this bias can be seen qualitatively
in the right column of Figure~\ref{fig_los_plot}.
Optimal redshift resolution requires choosing a 
filtering level which balances the effects of noise and bias,
and may require some form of bias correction.  In future work, we will
explore in detail the ways in which the SVD method allows
for a near optimal reconstruction of projected mass maps
and halo redshifts from data on galaxy shapes and photometric redshifts. 

\subsection{Scalability}
\label{Scalability} 

As we look forward to future surveys, it becomes important to consider methods 
that will scale upward with increasing survey volumes. Present weak lensing
surveys cover fields on the order of a few square degrees 
\citep[e.g.\ COSMOS,][]{Massey07}.  Future surveys will increase the field
size exponentially: up to $\sim\!\!20,000$ square degrees for LSST
\citep{LSST09}.  Though the flat-sky approximation used in this 
work is not appropriate for such large survey
areas, the weak lensing formalism can be modified to 
account for spherical geometry \citep[see, e.g.][]{Heavens03}. 

The main computational cost for both SVD and Wiener filtering is the
Fast Fourier Transform (FFT) required to implement the mapping from
$\gamma$ to $\kappa$.  For an $N \times N$ pixel field, the FFT algorithm 
performs in $\mathcal{O}[N\log N]$ in each dimension, 
meaning that the 2D FFT takes 
$\mathcal{O}[(N\log N)^2] \approx \mathcal{O}[N^2]$.  The Wiener
filter method, however, requires the inversion of a very large matrix 
using, for example, a conjugate-gradient method.  
The exact number of iterations
depends highly on the condition number of the matrix to be inverted;
STH09 finds that up to 150 iterations are required for this problem.
We find that \textit{each iteration} takes over 3 times longer than the 
entire SVD reconstruction. The net result is that both algorithms
scale nearly linearly with the area of the field (for constant pixel scale),
though the SVD estimator is computed up to 500 times faster 
than the Wiener filter. 

Extrapolating this scaling, the appropriately scaled 
SVD filter will allow reconstruction
of the entire $\sim\!\!20,000$ square-degree 
LSST field in a few hours on a single workstation, given
enough memory.  On the same computer, the Wiener-filter method would take 
over a month, depending on the amount and type of filtering and 
assuming that the required number of iterations stays constant with 
increasing field size. For the SVD-filtered reconstruction of this 
large field, the real challenge will not be computational time, 
but memory constraints: the complex shear
vector itself for such a field will require $\sim\!\!30$ GB of memory,
with the entire algorithm consuming approximately three times this.
The memory requirements for the Wiener filter will be comparable.
This is within reach of current high-end workstations as well as 
shared-memory parallel clusters.

\begin{table}
\centering
\caption{Masses and redshifts of halos in Figure~\ref{fig_many_halos}.}
\label{halo_table}
\begin{tabular}{lllll}
\hline
& $\theta_x$ & $\theta_y$ & $z$ & $M/M_\odot$ \\
\hline
  A & 37.5 & 44.9 & 0.60 & $7.2\times 10^{14}$ \\
  B & 67.1 & 70.5 & 0.47 & $6\times 10^{14}$ \\
  C & 46.9 & 106.5 & 0.63 & $5.5\times 10^{14}$ \\
  D & 108.9 & 94.3 & 0.63 & $5.4\times 10^{14}$ \\
  E & 97.9 & 63.6 & 0.39 & $4.9\times 10^{14}$ \\
  F & 102.4 & 84.8 & 0.70 & $3.8\times 10^{14}$ \\
  G & 77.0 & 49.6 & 0.58 & $3.2\times 10^{14}$ \\
  H & 52.0 & 48.5 & 0.36 & $3.2\times 10^{14}$ \\
  I & 72.6 & 45.6 & 0.78 & $2.9\times 10^{14}$ \\
  J & 68.6 & 64.5 & 0.68 & $2.5\times 10^{14}$ \\
  K & 8.6 & 34.5 & 0.32 & $2.3\times 10^{14}$ \\
  L & 10.5 & 49.5 & 0.51 & $2.3\times 10^{14}$ \\
  M & 99.4 & 56.5 & 0.22 & $2.3\times 10^{14}$ \\
  N & 21.7 & 53.1 & 0.76 & $2.3\times 10^{14}$ \\
  O & 31.6 & 102.1 & 0.69 & $2.2\times 10^{14}$ \\
  P & 69.7 & 33.2 & 0.39 & $2.2\times 10^{14}$ \\
\hline
\end{tabular}

\end{table}

\section{Conclusion}
\label{Conclusions}

We have presented a new method for producing tomographic maps of dark matter
through weak lensing, using truncation of singular values.  We have
tested and compared our method to the Wiener filter based
method of STH09, which is the first three-dimensional mass mapping
approach that is applicable to large area surveys. Our reconstruction
shares many of the aspects of the Wiener filter reconstruction,
in the sense that it massively reduces the noise inherent in the problem.
Our SVD method may be considered even more non-parametric
than the Wiener filter method, since it does not rely on any a priori 
assumptions of the statistical properties of the signal: all of the noise 
reduction is derived from the observed noise properties of the data.

The SVD framework allows a unique quantitative comparison between the
different filtering methods and filtering strengths.  Using the
coefficients of the weighted principal components contained in the SVD,
we have compared the three filtering methods, and have found that 
the radial Wiener filter of HK02 and SVD filter of this work are 
less-biased noise reduction techniques than the transverse Wiener filter
of STH09.  These authors have recently implemented the radial Weiner 
filter and obtain results consistent with our findings (P. Simon and 
A. Taylor, private communication).

The angular resolution of the SVD-reconstructed mass maps seems to 
be significantly better than that of the transverse Wiener filter method,
the method chosen in the STH09 analysis.
This allows for more robust separation of pairs of halos into two 
separate halos rather than blurring them into a single mass peak.  
We discuss how our reconstruction method provides a scheme for
optimizing the 3D reconstruction of projected mass maps by
balancing the goals of robustness of detecting specific structures
and improved redshift resolution. 

The SVD method can compute the three-dimensional mass maps
rapidly provided sufficient computational memory is available. 
This allows for the possibility of solving the full-sky tomographic
lensing inversion on the scale of hours, rather than months, which
makes it readily applicable to upcoming surveys. 

On the other hand, the redshift resolution with the SVD method is not
significantly  better than that of either Wiener filter method.  
This was a problem identified by STH09, and unfortunately
the SVD method does not significantly improve the situation. 
Our analysis of the noise characteristics of radial modes 
indicates that linear, non-parametric reconstruction methods are
fundamentally limited in this regard.

{\it Acknowledgments:} We are grateful to Patrick Simon and Andy
Taylor for numerous discussions and insights on the Wiener filter method, 
which helped improve our study.  We thank the anonymous referee for
several helpful suggestions.
Support for this research was provided by NASA Grant NNX07-AH07G.
We acknowledge partial support from NSF AST-0709394.

%%%%%%%%%%%%%%%%%%%%%%%%%%%%%%%%%%%%%%%%%%%%%%%%%%%%%%%%%%%%%%%%%%%%%%%%%%%
%             begin appendix

\appendix
\section{Efficient Implementation of the SVD Estimator}
\label{app}
As noted in Section~\ref{sing_val_formalism}, taking the SVD of the 
transformation matrix $\widetilde{M}_{\gamma\delta} \equiv 
\mathcal{N}_{\gamma\gamma}^{-1/2}M_{\gamma\delta}$
is not trivial for large fields.  This appendix will first give a rough
outline of the form of $M_{\gamma\delta}$, then describe our tensor 
decomposition method which enables quick calculation of the singular
value decomposition.  For a more thorough review of the lensing results, 
see e.g.~\citet{Bartelmann01}.

Our goal is to speed the computation of the SVD by writing \
$\widetilde{M}_{\gamma\delta}$
as a tensor product $\mymat{A} \otimes \mymat{B}$.  Here ``$\otimes$''
is the Kronecker product, defined such that, if $\mymat{A}$ is a matrix
of size $n \times m$, $B$ is a matrix of arbitrary size,
\begin{equation}
  \mymat{A}\otimes\mymat{B} \equiv \left(
  \begin{array}{cccc}
    A_{11}B & A_{12}B & \cdots & A_{1m}B \\
    A_{21}B & A_{22}B & \cdots & A_{2m}B \\
    \vdots  & \vdots & \ddots & \vdots  \\
    A_{n1}B & A_{n2}B & \cdots & A_{nm}B 
  \end{array}\right)
\end{equation}
In this case, the singular value decomposition
$A\otimes B = U_{AB}\Sigma_{AB}V^\dagger_{AB}$
satisfies
\begin{eqnarray}
  \label{AB_SVD}
  U_{AB} &=& U_A\otimes U_B \nonumber\\
  \Sigma_{AB} &=& \Sigma_A \otimes \Sigma_B\nonumber\\
  V_{AB} &=& V_A \otimes V_B
\end{eqnarray}
where $U_A\Sigma_AV^\dagger_A$ is the SVD of $A$, 
and   $U_B\Sigma_BV^\dagger_B$ is the SVD of $B$.
Decomposing $\widetilde{M}_{\gamma\delta}$ in this way can 
greatly speed the SVD computation.

\subsection{Angular and Line-of-Sight Transformations}
The transformation from shear to density, encoded in $M_{\gamma\delta}$,
consists of two steps: an angular integral relating shear $\gamma$ to
convergence $\kappa$, and a line-of-sight integral relating the convergence
$\kappa$ to the density contrast $\delta$.

The relationship between $\gamma$ and $\kappa$ is a convolution over
all angular scales,
\begin{equation}
  \label{gamma_integral}
  \gamma(\myvec\theta,z_s) \equiv \gamma_1 + i\gamma_2 = \int \dd^2\theta^\prime
  \ \mathcal{D}(\myvec\theta^\prime-\myvec\theta)\kappa(\myvec\theta^\prime,z_s),
\end{equation}
where $\mathcal{D}(\myvec\theta)$ 
is the Kaiser-Squires kernel \citep{Kaiser93}.  
This has a particularly simple form in Fourier space:
\begin{equation}
  \label{gamma_fourier}
  \hat\gamma(\myvec\ell,z_s) 
  = \frac{\ell_1 + i\ell_2}{\ell_1 - i\ell_2}\hat\kappa(\myvec\ell,z_s).
\end{equation}
where $\hat\gamma$ and $\hat\kappa$ are the Fourier transforms of $\gamma$
and $\kappa$ and $\myvec{\ell}\equiv(\ell_1,\ell_2)$ is the angular wavenumber.

The relationship between $\kappa$ and $\delta$ is an integral along each
line of sight:
\begin{equation}
  \label{kappa_integral}
  \kappa(\myvec\theta,z_s) = 
  \int_0^{z_s}\dd z\ W(z,z_s)\delta(\myvec\theta,z)
\end{equation}
where $W(z,z_s)$ is the lensing efficiency function at redshift $z$ 
for a source located at redshift $z_s$ 
(refer to STH09 for the form of this function).

Upon discretization of the quantities $\gamma$, $\kappa$, and $\delta$
(described in Section~\ref{LinearMapping}), 
the integrals in Equations~\ref{gamma_integral}-\ref{kappa_integral} 
become matrix operations.  The relationship between the data vectors
$\myvec{\gamma}$ and $\myvec{\kappa}$ can be written
\begin{equation}
  \label{P_gk}
  \myvec\gamma = [\mymat{P}_{\gamma\kappa} \otimes \mathbf{1}_s]\myvec\kappa 
  + \myvec{n}_\gamma
\end{equation}
where $\mathbf{1}_s$ is the $N_s \times N_s$ identity matrix and 
$\mymat{P}_{\gamma\kappa}$ is the matrix representing the linear 
transformation in Equations~\ref{gamma_integral}-\ref{gamma_fourier}.  
The quantity $[\mymat{P}_{\gamma\kappa} \otimes \mathbf{1}_s]$ 
simply denotes that $\mymat{P}_{\gamma\kappa}$ operates on each of the $N_s$ 
source-planes represented within the vector $\myvec\kappa$.
Similarly, the relationship between the vectors $\myvec{\kappa}$ and
$\myvec{\delta}$ can be written
\begin{equation}
  \label{Q_kd}
  \myvec\kappa = [\mathbf{1}_{xy} \otimes \mymat{Q}_{\kappa\delta}]\myvec\delta
\end{equation}
where $\mathbf{1}_{xy}$ is the $N_{xy} \times N_{xy}$ 
identity matrix, and the tensor product signifies that the operator 
$Q_{\kappa\delta}$ operates on each of the $N_{xy}$ lines-of-sight in
$\myvec\delta$.  $Q_{\kappa\delta}$ is the $N_s \times N_l$ matrix which
represents the discretized version of equation \ref{kappa_integral}.
Combining these representations allows us to decompose the matrix 
$\mymat{M}_{\gamma\delta}$ in Equation~\ref{M_gd} into a tensor product:
\begin{equation}
  \mymat{M}_{\gamma\delta} = 
  \mymat{P}_{\gamma\kappa} \otimes \mymat{Q}_{\kappa\delta}.
\end{equation}

\subsection{Tensor Decomposition of the Transformation}
We now make an approximation that the noise covariance 
$\mymat{\mathcal{N}}_{\gamma\gamma}$ can be written as a
tensor product between its angular part $\mymat{\mathcal{N}_P}$ 
and its line of sight part $\mymat{\mathcal{N}_Q}$:
\begin{equation}
  \label{noise_decomp}
  \mymat{\mathcal{N}}_{\gamma\gamma} 
  = \mymat{\mathcal{N}_P} \otimes \mymat{\mathcal{N}_Q}.
\end{equation}
Because shear measurement error comes primarily from shot noise, this 
approximation is equivalent to the statement that source galaxies are drawn 
from a single redshift distribution, with a different normalization along 
each line-of-sight.  For realistic data, this approximation will break down
as the size of the pixels becomes very small.  We will assume here for 
simplicity that the noise covariance is diagonal, but the following results
can be generalized for non-diagonal noise.  
Using this noise covariance approximation, we can compute the 
SVDs of the components of $\widetilde{M}_{\gamma\delta}$:
\begin{eqnarray}
  \mymat{U}_P\mymat{\Sigma}_P\mymat{V}_P^\dagger = \mathcal{N}_P^{-1/2} \mymat{P}_{\gamma\kappa}\nonumber\\
  \mymat{U}_Q\mymat{\Sigma}_Q\mymat{V}_Q^\dagger = \mathcal{N}_Q^{-1/2} \mymat{Q}_{\kappa\delta}
\end{eqnarray}

In practice the SVD of the matrix $\mymat{P}_{\gamma\kappa}$ 
need not be computed explicitly.  
$\mymat{P}_{\gamma\kappa}$ encodes the discrete linear operation expressed
by Equations~\ref{gamma_integral}-\ref{gamma_fourier}: 
as pointed out by STH09, in the large-field limit $P_{\gamma\kappa}$ 
can be equivalently computed in either real or Fourier space.
Thus to operate with $P_{\gamma\kappa}$ on a shear vector, 
we first take the 2D Fast Fourier Transform (FFT) of each
source-plane, multiply by the kernel $(\ell_1+i\ell_2)/(\ell_1-i\ell_2)$,
then take the inverse FFT of the result.  This is orders-of-magnitude
faster than a discrete implementation of the real-space convolution.
Furthermore, the conjugate transpose of this operation can be computed
by transforming $\ell \to -\ell^*$, so that
\begin{equation}
  \mymat{P}_{\gamma\kappa}^\dagger\mymat{P}_{\gamma\kappa} = \mymat{I}
\end{equation}
and we see that $P_{\gamma\kappa}$ is unitary in the wide-field limit.  This
fact, along with the tensor product properties of the SVD, allows us to
write $\widetilde{M}_{\gamma\delta} = U\Sigma V^\dagger$ where
\begin{eqnarray}
  U &\approx& \mathbf{1}_{xy} \otimes U_Q \nonumber\\
  \Sigma &\approx& \mathcal{N}_P^{-1/2} \otimes \Sigma_Q \nonumber\\
  V^\dagger &\approx& P_{\gamma\kappa} \otimes  V_Q^\dagger
\end{eqnarray}
The only explicit SVD we need to calculate is that of 
$\mathcal{N}_Q^{-1/2}\mymat{Q}_{\kappa\delta}$,
which is trivial in cases of interest.  
The two approximations we have made are the
applicability of the Fourier-space form of the $\gamma\to\kappa$ mapping
(Eqn.~\ref{gamma_fourier}), and the tensor
decomposition of the noise covariance (Eqn.~\ref{noise_decomp}).

%%%%%%%%%%%%%%%%%%%%%%%%%%%%%%%%%%%%%%%%%%%%%%%%%%%%%%%%%%%%%%%%%%%%%%%%%%%
%             begin bibliography
\bibliography{letter}

\begin{thebibliography}{16}
\expandafter\ifx\csname natexlab\endcsname\relax\def\natexlab#1{#1}\fi

\bibitem[{Aitken(1934)}]{Aitken34}
Aitken, A. 1934, Proc. R. Soc. Edinb, 55, 42

\bibitem[{{Bacon} \& {Taylor}(2003)}]{Bacon03}
{Bacon}, D.~J., \& {Taylor}, A.~N. 2003, \mnras, 344, 1307

\bibitem[{{Bartelmann} \& {Schneider}(2001)}]{Bartelmann01}
{Bartelmann}, M., \& {Schneider}, P. 2001, \physrep, 340, 291

\bibitem[{{Clowe} {et~al.}(2006){Clowe}, {Brada{\v c}}, {Gonzalez},
  {Markevitch}, {Randall}, {Jones}, \& {Zaritsky}}]{Clowe06}
{Clowe}, D., {Brada{\v c}}, M., {Gonzalez}, A.~H., {Markevitch}, M., {Randall},
  S.~W., {Jones}, C., \& {Zaritsky}, D. 2006, \apjl, 648, L109

\bibitem[{{de Putter} \& {White}(2005)}]{dePutter05}
{de Putter}, R., \& {White}, M. 2005, New A, 10, 676

\bibitem[{{Heavens}(2003)}]{Heavens03}
{Heavens}, A. 2003, \mnras, 343, 1327

\bibitem[{{Hoekstra}(2003)}]{Hoekstra03}
{Hoekstra}, H. 2003, \mnras, 339, 1155

\bibitem[{{Hu} \& {Keeton}(2002)}]{Hu02}
{Hu}, W., \& {Keeton}, C.~R. 2002, \prd, 66, 063506

\bibitem[{{Kaiser} \& {Squires}(1993)}]{Kaiser93}
{Kaiser}, N., \& {Squires}, G. 1993, \apj, 404, 441

\bibitem[{{LSST Science Collaborations} {et~al.}(2009){LSST Science
  Collaborations}, {Abell}, {Allison}, {Anderson}, {Andrew}, {Angel}, {Armus},
  {Arnett}, {Asztalos}, {Axelrod}, {Bailey}, {Ballantyne}, {Bankert},
  {Barkhouse}, {Barr}, {Barrientos}, {Barth}, {Bartlett}, {Becker}, {Becla},
  {Beers}, {Bernstein}, {Biswas}, {Blanton}, {Bloom}, {Bochanski}, {Boeshaar},
  {Borne}, {Bradac}, {Brandt}, {Bridge}, {Brown}, {Brunner}, {Bullock},
  {Burgasser}, {Burge}, {Burke}, {Cargile}, {Chandrasekharan}, {Chartas},
  {Chesley}, {Chu}, {Cinabro}, {Claire}, {Claver}, {Clowe}, {Connolly}, {Cook},
  {Cooke}, {Cooray}, {Covey}, {Culliton}, {de Jong}, {de Vries}, {Debattista},
  {Delgado}, {Dell'Antonio}, {Dhital}, {Di Stefano}, {Dickinson}, {Dilday},
  {Djorgovski}, {Dobler}, {Donalek}, {Dubois-Felsmann}, {Durech},
  {Eliasdottir}, {Eracleous}, {Eyer}, {Falco}, {Fan}, {Fassnacht}, {Ferguson},
  {Fernandez}, {Fields}, {Finkbeiner}, {Figueroa}, {Fox}, {Francke}, {Frank},
  {Frieman}, {Fromenteau}, {Furqan}, {Galaz}, {Gal-Yam}, {Garnavich},
  {Gawiser}, {Geary}, {Gee}, {Gibson}, {Gilmore}, {Grace}, {Green}, {Gressler},
  {Grillmair}, {Habib}, {Haggerty}, {Hamuy}, {Harris}, {Hawley}, {Heavens},
  {Hebb}, {Henry}, {Hileman}, {Hilton}, {Hoadley}, {Holberg}, {Holman},
  {Howell}, {Infante}, {Ivezic}, {Jacoby}, {Jain}, {R}, {Jedicke}, {Jee},
  {Garrett Jernigan}, {Jha}, {Johnston}, {Jones}, {Juric}, {Kaasalainen},
  {Styliani}, {Kafka}, {Kahn}, {Kaib}, {Kalirai}, {Kantor}, {Kasliwal},
  {Keeton}, {Kessler}, {Knezevic}, {Kowalski}, {Krabbendam}, {Krughoff},
  {Kulkarni}, {Kuhlman}, {Lacy}, {Lepine}, {Liang}, {Lien}, {Lira}, {Long},
  {Lorenz}, {Lotz}, {Lupton}, {Lutz}, {Macri}, {Mahabal}, {Mandelbaum},
  {Marshall}, {May}, {McGehee}, {Meadows}, {Meert}, {Milani}, {Miller},
  {Miller}, {Mills}, {Minniti}, {Monet}, {Mukadam}, {Nakar}, {Neill}, {Newman},
  {Nikolaev}, {Nordby}, {O'Connor}, {Oguri}, {Oliver}, {Olivier}, {Olsen},
  {Olsen}, {Olszewski}, {Oluseyi}, {Padilla}, {Parker}, {Pepper}, {Peterson},
  {Petry}, {Pinto}, {Pizagno}, {Popescu}, {Prsa}, {Radcka}, {Raddick},
  {Rasmussen}, {Rau}, {Rho}, {Rhoads}, {Richards}, {Ridgway}, {Robertson},
  {Roskar}, {Saha}, {Sarajedini}, {Scannapieco}, {Schalk}, {Schindler},
  {Schmidt}, {Schmidt}, {Schneider}, {Schumacher}, {Scranton}, {Sebag},
  {Seppala}, {Shemmer}, {Simon}, {Sivertz}, {Smith}, {Allyn Smith}, {Smith},
  {Spitz}, {Stanford}, {Stassun}, {Strader}, {Strauss}, {Stubbs}, {Sweeney},
  {Szalay}, {Szkody}, {Takada}, {Thorman}, {Trilling}, {Trimble}, {Tyson}, {Van
  Berg}, {Vanden Berk}, {VanderPlas}, {Verde}, {Vrsnak}, {Walkowicz},
  {Wandelt}, {Wang}, {Wang}, {Warner}, {Wechsler}, {West}, {Wiecha},
  {Williams}, {Willman}, {Wittman}, {Wolff}, {Wood-Vasey}, {Wozniak}, {Young},
  {Zentner}, \& {Zhan}}]{LSST09}
{LSST Science Collaborations} {et~al.} 2009, ArXiv e-prints

\bibitem[{{Massey} {et~al.}(2007){Massey}, {Rhodes}, {Leauthaud}, {Capak},
  {Ellis}, {Koekemoer}, {R{\'e}fr{\'e}gier}, {Scoville}, {Taylor}, {Albert},
  {Berg{\'e}}, {Heymans}, {Johnston}, {Kneib}, {Mellier}, {Mobasher},
  {Semboloni}, {Shopbell}, {Tasca}, \& {Van Waerbeke}}]{Massey07}
{Massey}, R., {et~al.} 2007, \apjs, 172, 239

\bibitem[{{Navarro} {et~al.}(1997){Navarro}, {Frenk}, \& {White}}]{NFW97}
{Navarro}, J.~F., {Frenk}, C.~S., \& {White}, S.~D.~M. 1997, \apj, 490, 493

\bibitem[{{Rines} {et~al.}(2007){Rines}, {Diaferio}, \& {Natarajan}}]{Rines07}
{Rines}, K., {Diaferio}, A., \& {Natarajan}, P. 2007, \apj, 657, 183

\bibitem[{{Simon} {et~al.}(2009){Simon}, {Taylor}, \& {Hartlap}}]{Simon09}
{Simon}, P., {Taylor}, A.~N., \& {Hartlap}, J. 2009, \mnras, 399, 48

\bibitem[{{Takada} \& {Jain}(2003)}]{Takada03}
{Takada}, M., \& {Jain}, B. 2003, \mnras, 344, 857

\bibitem[{{Taylor}(2001)}]{Taylor01}
{Taylor}, A.~N. 2001, ArXiv Astrophysics e-prints

\end{thebibliography}

\end{document}